\documentclass[useAMS,usenatbib]{mn2e}
\usepackage{rotating}
\usepackage{graphicx}
\usepackage{epsf}
\usepackage{aas_macros}
\newcommand{\kms}{{\rm {km\, s^{-1}}}}

\newcommand{\msun}{M_{\odot }}
\newcommand{\Mo}{\msun}
\newcommand{\Ro}{R_{\odot }}

\newcommand{\ud}{{\rm d}}

\newcommand{\BSE}{{\sc BSE}}

\title[Spatial offsets of short-duration GRBs]
{
Implications for the origin of short gamma-ray bursts from their
observed positions around their host galaxies
}

\author[R.P.~Church et al.]{Ross P.~Church$^{1}$\thanks{email: ross@astro.lu.se},
Andrew J. Levan$^{2,1}$, Melvyn B. Davies$^{1}$ and Nial Tanvir$^{3}$\\
$^{1}$Lund Observatory, Box 43, SE--221 00, Lund, Sweden.\\
$^{2}$Department of Physics, University of Warwick, Coventry, CV4 7AL \\
$^{3}$Department of Physics and Astronomy, University of Leicester,
Leicester, LE1~7RH, UK. \\
}

\begin{document}

\date{Accepted 2011 January 4. Received 2010 November 17; in original form 2010
 January 7.}

\pagerange{\pageref{firstpage}--\pageref{lastpage}} \pubyear{2011}

\maketitle

\label{firstpage}

\begin{abstract}
We present the observed offsets of short-duration gamma-ray bursts (SGRBs) from
their putative host galaxies and compare them to the expected distributions of
merging compact object binaries, given the observed properties of the hosts.
We find that for all but one burst in our sample the offsets are consistent
with this model.  For the case of bursts with massive elliptical host galaxies,
the circular velocities of the hosts' haloes exceed the natal velocities of
almost all our compact object binaries.  Hence the extents of the predicted
offset distributions for elliptical galaxies are determined largely by their
spatial extents.  In contrast, for spiral hosts the galactic rotation
velocities are smaller than typical binary natal velocities and the predicted
burst offset distributions are more extended than the galaxies.  

One SGRB, 060502B, apparently has a large radial offset that is inconsistent
with an origin in a merging galactic compact binary.  Although it is plausible
that the host of GRB~060502B is mis-identified, our results show that the large
offset is compatible with a scenario where at least a few per cent of SGRBs are
created by the merger of compact binaries that form dynamically in globular
clusters.

\end{abstract}


\section{Introduction}
The merger of a compact binary consisting of neutron stars (NSs) or black holes
(BHs) is a prime model for the origin of the short-duration gamma-ray bursts
(SGRBs). Compact binaries may be formed by numerous channels including those
which are formed in primordial binaries \citep{bel2002b}, and those which are
formed dynamically in dense cluster cores
\citep[e.g.][]{davies1995,grindlay2005}. Once formed the binary loses energy
and angular momentum from its orbit in the form of gravitational radiation, and
the compact objects spiral together and eventually merge.  The merger time is
highly sensitive to both separation and eccentricity and thus has an extremely
broad distribution.

Evidence for the origin of SGRBs in the final coalescence of such systems
potentially comes both from the host galaxy types \citep[e.g.][]{zheng2007} and
the measured offsets of SGRBs from their hosts
\citep[e.g.][]{bloom99,bel2006,troja08}.  The kicks imparted to double compact
object (DCO) systems on formation, both in terms of dynamical kicks to the
binary and natal kicks to neutron stars, endow DCO binaries with spatial
velocities of up to several hundred $\kms$. This means that they may merge far
from their birth sites, even outside their parent galaxies. Equally, if the
inspiral takes $>10^8\,{\rm years}$, then the host galaxy may no longer exhibit
significant ongoing star formation at the time of merger.

Observations of SGRBs obtained by {\it Swift} so far offer broad support for
this picture. Several of the host galaxies observed are old
\citep{gehrels2005,berger2005,barthelmy2005,bloom2006}, while only a small
fraction are starbursts.  However, the sample has been building up slowly, and
we are only now approaching a stage where it is possible to compare the observed
offsets with predictions for DCO binaries.  We build on previous work that
modelled the offset distributions of DCO mergers \citep[e.g.][]{bloom99,fryer99}
by using the observed properties of the host galaxies to predict offset
distributions on a host-by-host basis.

\begin{table*}
\begin{center}
\begin{tabular}{llllllllllllll}
\hline
GRB    & Satellite & $t_{90}$  & Opt & Host Mag. & Host $M_B$   & z& Type & $R_{\rm e}$ & $R_{\rm proj}$& $R_{\rm err}$& References\\
       &           &  s        &     &           &              &  &      & kpc           &  kpc      & kpc      &\\
\hline
050509B & {\it Swift}  & 0.04 & No  & 16.75 $\pm$ 0.05 & $-23.25$ & 0.225 & E & 20.98  & 63.7  & 12.1 & (a), (b)\\
050709  & {\it HETE-2} & 0.07 & Yes & 21.05 $\pm$ 0.07 & $-18.19$ & 0.161 & S/I & 1.75 & 3.55  & 0.27 & (c), (d), (b)\\
050724  & {\it Swift}  & 3.0  & Yes & 18.19 $\pm$ 0.03 & $-22.11$ & 0.257 & E & 4.00   & 2.54  & 0.08 & (e), (b)\\
051221A & {\it Swift}  & 1.4  & Yes & 21.81 $\pm$ 0.09 & $-20.20$ & 0.546 & S & 2.17   & 1.53  & 0.31 & (f), (g)\\
060502B & {\it Swift}  & 0.05 & No  & 18.71 $\pm$ 0.01 & $-21.84$ & 0.287 & E & 10.5   & 73    & 13   & (h)\\
060801  & {\it Swift}  & 0.5  & No  & 22.97 $\pm$ 0.11 & $-20.64$ & 1.130 & S & -      & 19.7  & 14.0 & (i), (j)\\
061006  & {\it Swift}  & 0.4  & Yes & 22.65 $\pm$ 0.09 & $-18.86$ & 0.438 & S & 3.67   & 1.44  & 0.29 & (b)\\
061201  & {\it Swift}  & 0.8  & Yes & 19.65 $\pm$ 0.10 & $-18.76$ & 0.111 & S & 1.8    & 33.9  & 0.4  & (i), (j)\\
061210  & {\it Swift}  & 0.2  & No  & 21.00 $\pm$ 0.02 & $-20.36$ & 0.410 & S & -      & 10.7  & 6.9  & (i), (j)\\
061217  & {\it Swift}  & 0.3  & No  & 23.33 $\pm$ 0.07 & $-19.61$ & 0.827 & S & -      & 55    & 20   & (i), (j)\\
070429B & {\it Swift}  & 0.5  & Yes & 23.22 $\pm$ 0.10 & -19.91 & 0.902 & S & -      & 4.7   & 4.7  & (k)\\
070714B & {\it Swift}  & 2.0  & Yes & 24.92 $\pm$ 0.23 & -18.26 & 0.923 & S & -      & 3.08  & 0.47 & (l)\\
070724A & {\it Swift}  & 0.4  & Yes & 20.53 $\pm$ 0.03 & -21.07 & 0.457 & E & -      & 4.76  & 0.06 & (m)\\
070809  & {\it Swift}  & 1.3  & Yes & 21.7  $\pm$ 0.3  & -18.23 & 0.219 & S & -      & 19.61 & 1.9  & (n)\\
071227  & {\it Swift}  & 1.8  & Yes & 20.54 $\pm$ 0.03 & -20.73 & 0.394 & S & -      & 16.1  & 0.2  & (o)\\
080905A & {\it Swift}  & 1.0  & Yes & 18.0  $\pm$ 0.5  & -20.63 & 0.122 & S & -      & 18.11 & 0.42 & (p)\\
\hline
\end{tabular}
\end{center}

\caption{
Properties of SGRBs with known X-ray, optical or radio counterparts.  Data is
taken from the literature sources indicated other than for a small number of
the offsets which we have re-measured in described in Section~2.  Columns list
in order: the burst identifier, the satellite which detected the burst, the
duration over which 90\% of the total fluence was seen ($t_{90}$), whether an
optical counterpart was detected, the host apparent $R$-band magnitude and
error, the host absolute $B$-band magnitude, the redshift, the host type
(spiral/elliptical), the host effective radius $R_{\rm e}$, the offset $R_{\rm
proj}$ and the error on the offset $R_{\rm err}$.  The duration, $t_{90}$, is
measured in the 15-350 keV range for the {\it Swift} bursts and in the 30-400
keV range for the {\it HETE-2} burst.  The absolute magnitudes have been
calculated assuming a source with a flat spectrum in $F_{\nu}$.  All offset
errors are one-sigma; 90\% confidence limits in the literature have been
converted assuming a Gaussian PSF.  References: (a) \citet{gehrels2005}; (b)
\citet{fong09}; (c) \citet{fox05}; (d) \citet{villasenor05}; (e)
\citet{berger2005}; (f) \citet{soderberg06}; (g) this work; (h)
\citet{bloom2007}; (i) \citet{bergerhiz}; (j) \citet{troja08}; (k)
\citet{Cenko09}; (l) \citet{graham09}; (m) \citet{berger09}; (n) Perley et al.
(GCN 7889); (o) \citet{davanzo09}; (p) Rowlinson et al. in prep.
} 
\label{tab:bursts}
\end{table*}%

In this paper we take the observed sample of SGRBs that have identified hosts
and model the production and galactic trajectories of DCO binaries inside those
hosts.  In Section 2 we review the sample of bursts; we describe our models in
Section 3, discussing the results and possible selection effects in Section 4.
The alternative possibility of DCO binaries forming dynamically in the cores of
globular clusters is discussed in Section 5.

\section{The bursts and their host galaxies}
\label{sect:burstshosts}
From the set of observed short GRBs we have selected the bursts with positions
localised through observations of their X-ray and, where possible, optical
afterglows.  Furthermore we have discarded bursts without identified host
galaxies or for which there are not robust redshift measurements, since
knowledge of the redshift is necessary to construct model galactic potentials.
For several bursts we have re-analysed available data to perform relative
astrometry at times when the afterglow was bright, mapping the locations of the
burst onto the hosts utilising the {\tt geomap} task within IRAF and a number
of point sources within the field. 
For other bursts we have used offsets available in the literature.
Our final sample of 16 bursts is listed in Table~\ref{tab:bursts}.

In order to obtain dynamical models for the bursts' host galaxies we split the
sample into elliptical and spiral hosts.  In both cases we adopt the
the logarithmic profile of \citet{thomas09} to model the host galaxy's dark
halo,
\begin{equation}
\rho = \frac{v_{\rm h}}{4\pi G}\frac{3r_{\rm h}^2+r^2}{(r_{\rm h}^2+r^2)^2},
\end{equation}
where $r_{\rm h}$ is the core radius of the halo and $v_{\rm h}$ its circular
velocity at infinity.  This profile is constructed such that the circular
velocity approaches $v_{\rm h}$ asymptotically at infinity.  It should be noted
that as elliptical galaxies do not have discs in their case the circular velocity
is not directly measurable, but it is useful to parameterise the potential in
the same way.  Initially we place
the binaries in an exponential disc, chosen so that half the stars are within
the effective radius.  For elliptical hosts we use the scaling relations of
\citet{thomas09} to obtain the properties of the galaxy's halo from its
blue-band magnitude.  The effective radius $R_{\rm e}$ is taken from
\citet{gerhard01} as cited by \citet{thomas09}.  The properties of spiral
galaxies are taken from \citet{Kormendy04}.  The complete set of models that we
used are given in Table~\ref{tab:models}.  We take $R_{\rm e}$ to be the
half-light radius of the galaxy.  A comparison between the observed and
predicted $R_{\rm e}$ shows that there is considerable scatter, as would be
expected from observations of galaxy properties \citep[see e.g. figure 3
of][]{thomas09}.  Where a galaxy has an observed $R_{\rm e}$ we compute models
using both the value predicted by the formula and the measured value, and use
the observational value in preference (for example, in the cumulative offset distributions of Figures~\ref{fig:cumProb060502B}~and~\ref{fig:cumProb061201}).

\begin{table}
\begin{center}
\begin{tabular}{llllllll}
\hline
GRB       & Type & $v_{\rm h}$        & $r_{\rm h}$ &  $R_{\rm e}$   & $R_{\rm e}^{\rm obs}$ & $M_{\rm h}$ \\
          &      & ${\rm km\,s^{-1}}$ & ${\rm kpc}$ &  ${\rm kpc}$   & $ {\rm kpc}$          & $10^{11}\,\Mo$ \\
\hline
%GRB     T       v_h      r_h      r_eff  r_eff(obs)  M_enc
%               km/s      kpc      kpc    kpc         10^11 Msun
050509B &E&   663.74 &   46.34 &  24.54 &  20.98 & 205 \\
050709  &S&   109.63 &    7.90 &   0.91 &  1.75  &   0.41 \\
050724  &E&   532.40 &   23.92 &   8.41 &  4.0   &  19.4 \\
051221A &S&   157.30 &   15.66 &   2.42 &  2.17  &   0.82 \\
060502B &E&   505.31 &   20.45 &   6.52 &  10.5  &  60.1 \\
060801  &S&   170.23 &   18.20 &   3.00 &        &   1.48 \\
061006  &S&   123.65 &    9.92 &   1.26 &  3.67  &   1.22 \\
061201  &S&   121.45 &    9.59 &   1.20 &  1.80  &   0.48 \\
061210  &S&   161.88 &   16.54 &   2.62 &        &   1.14 \\
061217  &S&   141.48 &   12.81 &   1.81 &        &   0.56 \\
070429B &S&   149.31 &   14.19 &   2.10 &        &   0.75 \\
070714B &S&   111.02 &    8.90 &   0.94 &        &   0.14 \\
070724A &E&   435.39 &   13.08 &   3.17 &        &  11.9 \\
070809  &S&   110.42 &    8.00 &   0.92 &        &   0.15 \\
071227  &S&   173.00 &   18.77 &   3.13 &        &   1.60 \\
080905A &S&   169.92 &   18.14 &   2.98 &        &   1.46 \\
%090510  &S&   133.82 &   11.53 &   1.56 &        \\

\hline
\end{tabular}
\end{center}
\caption{Parameters of the galactic models that we used.  Columns list the GRB,
whether the host is spiral or elliptical, the asymptotic rotational velocity of
the halo model $v_{\rm h}$, the halo core radius $r_{\rm h}$, the predicted
effective radius $R_{\rm e}$ and, where one is available, the observed effective
radius $R_{\rm e}^{\rm obs}$.  The final column is the mass enclosed within a
spherical radius of $10\,R_{\rm e}$, which acts an an indicative halo mass.}
\label{tab:models}
\end{table}

\section{Population synthesis and galactic trajectories}
To model the observed DCO populations we conduct a population synthesis of
NS--NS and BH--NS binaries, utilising the rapid binary population synthesis code
\BSE{} \citep{hurley2002}.  We have modified \BSE{} following \citet{bel2002b} and
\citet{bel2008} to include a more realistic prescription for compact object
masses, hypercritical accretion during common envelope evolution, and delayed
dynamical instability in mass transfer from helium stars in binaries with a
large mass ratio.  More details of our population synthesis code, and the
constraints which can be placed on DCO formation will be presented in a later
paper (Church et al.~in prep.).  For reasons of simplicity we distribute stellar
masses according to the \citet{kroupa1993} IMF, selecting both stars
independently from the IMF and only considering stars with masses greater than
$3\,\Mo$.  Binary semi-major axes are chosen from a distribution flat in $\log
a$ between 1 and $10^4\,\Ro$.  Preliminary simulations showed that binaries
wider than this do not contribute to the DCO population.  The initial
eccentricity of the binaries has little effect on their evolution and is set
equal to 0.1 for all binaries.

\subsection{Neutron star kicks}
For the purpose of modelling the galactic offsets of SGRBs the most significant
uncertainty is the natal kick imparted to neutron stars at their formation.  In
this paper we utilise two distributions of neutron star natal kicks: the bimodal
kick distribution of \citet{acc02}, henceforth ACC02, and the Maxwellian
distribution recommended by \citet{Dewi2005}, henceforth DPP05.  

The ACC02 distribution leads to strong kicks, with typical magnitudes of several
hundred $\kms$, and is derived from observations of isolated pulsars.  The DPP05
distribution yields much smaller kicks -- the dispersion is only $20\,\kms$ --
and is derived from constraints placed on the space velocity of the second
neutron star formed in a binary by the relationship between its spin and orbital
eccentricity.  We also computed models that utilised the distribution of
\citet{hansen1997}, but this produced results that were, for our purposes, very
similar to those obtained with the ACC02 distribution.

We specify the natal kick distributions for the two objects that form in each
DCO separately, considering three combinations of kick distributions.  In the
first scenario both compact objects receive an ACC02 (strong) kick.
Alternatively, the first compact object to form may receive an ACC02 (strong)
kick and the second a DPP05 (weak) kick. Finally both objects may receive a
DPP05 (weak) kick.  As the evidence for weak kicks relies on the properties of
the second compact object we do not consider the case where the first kick is
weak and the second kick is strong.

\begin{figure}
   \includegraphics[width=.75\columnwidth]{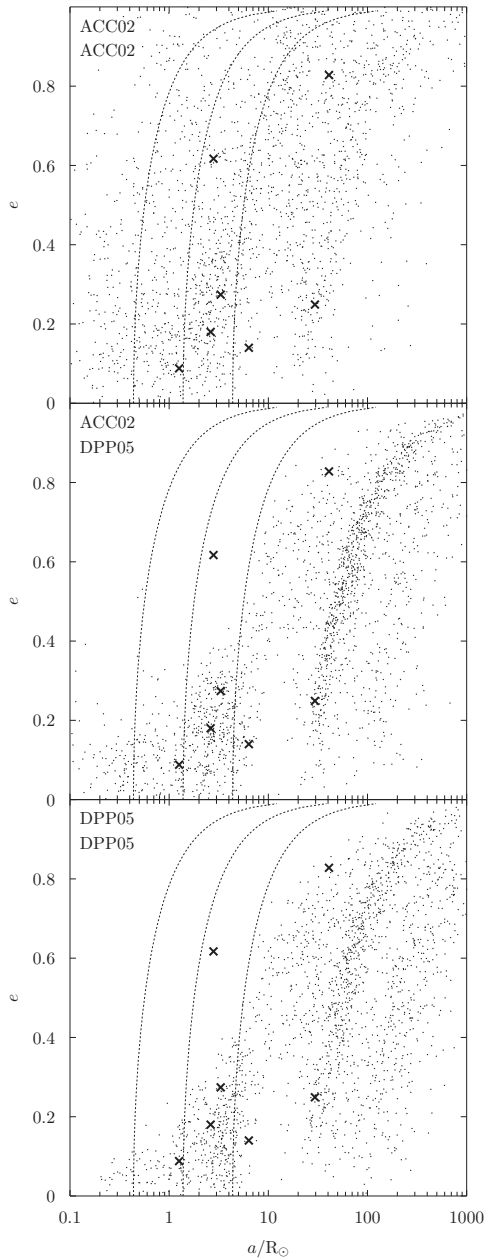}\\
   \caption{{\rm Dots show the eccentricity $e$ as a function of semi-major axis $a$
   for our synthesised populations of NS--NS binaries, plotted in the location
   in which they form.  To clarify the plots only 2000 binaries are plotted
   in each case.  Crosses show the properties of the observed NS--NS binaries.
   Lines of iso-merger-time of $10^6$, $10^8$ and $10^{10}\,{\rm yr}$ are also
   plotted.  The three panels show different combinations of kick distributions.
   Top: ACC02 kick for both NSs. Middle: ACC02 kick for the first NS and DPP05
   kick for the second NS.  Bottom: DPP05 kick for both NSs.}}
   \label{fig:avse}
\end{figure}

A powerful test of the neutron star kick distributions is the comparison of the
synthesised NS--NS binary populations that they yield with the properties of
observed NS--NS binaries.  In particular, the kicks very strongly affect the
semi-major axis $a$ and eccentricity $e$ of a given DCO binary.  The ability of
a kick distribution to produce the observed NS--NS binaries in the $a,e$ plane
is a stringent test of its correctness.  We present the $a,e$ distribution of
all the NS--NS binaries formed in our simulations in Figure~\ref{fig:avse}.

There is little evidence as to whether a kick is imparted to black holes at
formation \citep{bel2008}, though for the Galactic black-hole binary Cygnus X--1
there is evidence that there was no substantial natal kick
\citep{2003Sci...300.1119M}.  A significant kick would require the asymmetric
ejection of a large fraction of the progenitor's envelope, rather than its
accretion onto the black hole.  The larger mass of a typical black hole
compared to a neutron star increases the quantity of mass that must be ejected
to produce a significant kick.  Hence we choose not to apply a natal kick to
black holes upon formation.  Where a black hole forms via the
accretion-induced collapse of a neutron star we of course apply a kick to that
neutron star at its birth.  We have tested the effect of omitting black hole
kicks by synthesising a population of BH--NS binaries where a kick following
the ACC02 distribution is applied, reduced according to the ratio of the typical
neutron star mass to the nascent black hole mass.  This is equivalent to a natal
kick with the same distribution of impulses as a neutron star kick.  We found
that the space velocity distribution of the binaries that we produced was
largely unaffected.  This is unsurprising as such binaries are typically rather
massive and hence strongly bound at the time of the kick and, as our results
show, the properties of our sample of DCO binaries is relatively insensitive to
the natal kick chosen for the first supernova.

In addition to the natal kicks each supernova causes an impulse to the binary by
the rapid loss of a large quantity of mass.  As this mass escapes
anisotropically it takes away a significant net momentum.  This effect is also
taken into account in our simulations and accounts for the fact that, even when
two small kicks are utilised, a DCO binary may have a large spatial velocity
after formation.

\begin{figure}
   \includegraphics[width=.75\columnwidth]{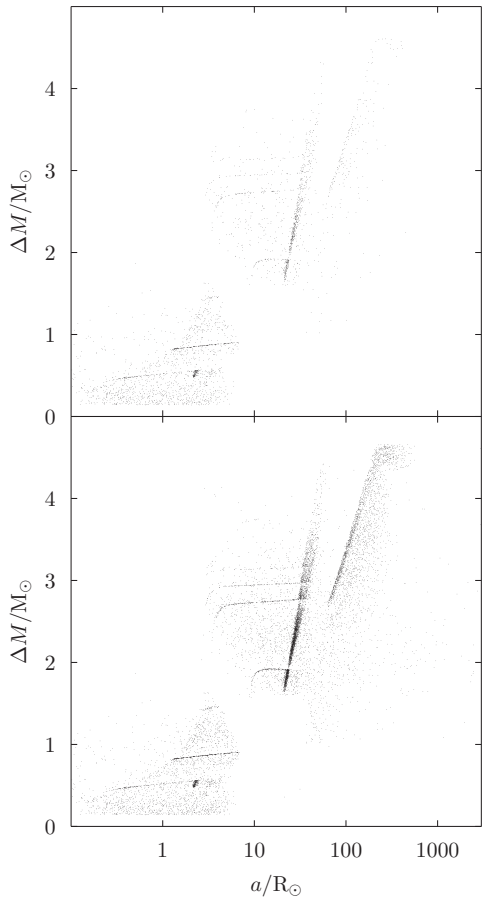}\\
   \caption{Dots show the mass lost in the final supernova, $\Delta M$, as a function of
   pre-supernova semi-major axis $a$ for the binaries that potentially form a
   NS--NS binary.  The bottom panel shows all binaries that are bound prior to
   the final supernova, the upper panel only the subset that remain bound
   subsequent to the final supernova.  The ACC02 kick distribution was used for
   both kicks.}
   \label{fig:avsdmNSNS}
\end{figure}

\begin{figure}
   \includegraphics[width=.75\columnwidth]{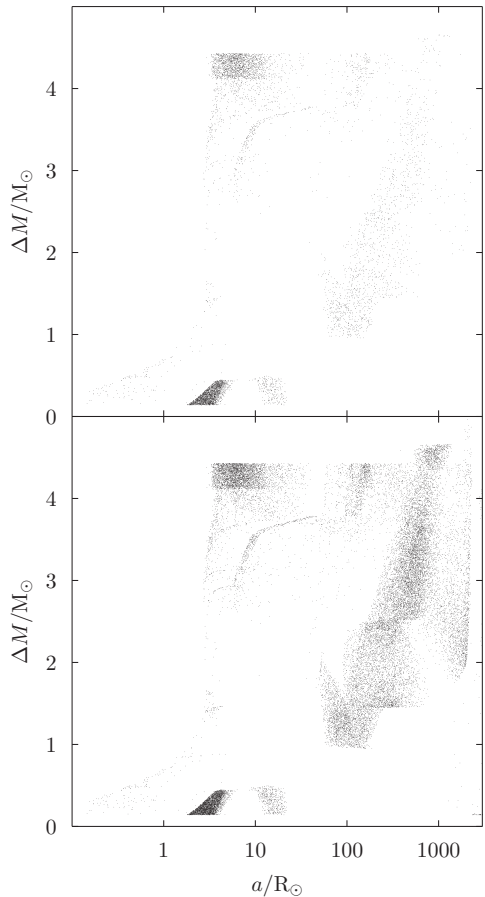}\\
   \caption{As in Figure~\ref{fig:avsdmNSNS} but for BH--NS progenitors.  Dots
   show the mass lost in the final supernova $\Delta M$ as a function of pre-supernova
   semi-major axis $a$ for the binaries that potentially form a NS--NS binary.
   The bottom panel shows all binaries that are bound prior to the final
   supernova, the upper panel only the subset that remain bound subsequent to
   the final supernova.  The ACC02 kick distribution was used for both kicks.}
   \label{fig:avsdmBHNS}
\end{figure}

We plot the mass lost in the final supernova as a function of the orbital
separation immediately prior to the supernova for all potential DCO
progenitors in Figures~\ref{fig:avsdmNSNS} (NS--NS binaries) and \ref{fig:avsdmBHNS}
(BH--NS binaries).  In each case the upper panel shows binaries that are bound
after the supernova and natal kick, whilst the lower panel includes systems
that unbind at the final supernova and hence form two isolated compact objects.
In both cases the natal kick
distribution is that which we prefer for the subsequent analysis, the ACC02
distribution (see Section~\ref{sect:discuss:pop}).  It is apparent in both cases
that the pre-supernova binaries fall into two groups, one with large separations
and one with small separations.  The large separation group tend to also have
larger mass-loss, as they have not gone through a recent episode of
common-envelope evolution and hence the progenitor of the second compact object
retains a larger fraction of its envelope.  These wider binaries are usually
disrupted by the final supernova, unless a chance alignment of natal kick causes
them to remain bound.  In the case where they do produce a bound binary it is
usually wide enough that it does not merge under the effects of gravitational
wave radiation within the Hubble time\footnote{The Hubble time is taken to be
$14\,{\rm Gyr}$ throughout.} and hence is not significant for our calculation.
The majority of systems that go on to merge and potentially form a gamma-ray
burst come from the population of closer binaries consisting of a compact object
and a helium star.

\begin{figure}
   \includegraphics[width=.75\columnwidth]{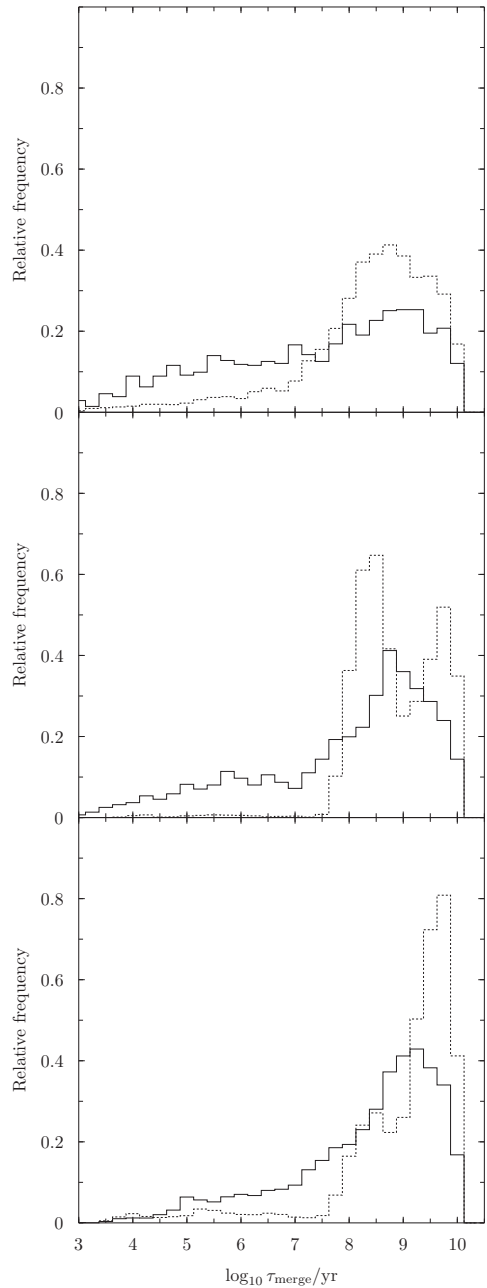}\\
   \caption{The distribution of merger times of the NS--NS and BH--NS binaries
   produced by our population synthesis calculations.  Solid lines show NS--NS
   binaries, dotted lines BH--NS binaries.  In the top panel the ACC02 kick
   distribution was used for both kicks, in the middle panel the ACC02
   distribution was used for the first kick and the DPP05 distribution for the
   second kick, and in the bottom panel the DPP05 distribution was used for both
   kicks.  Only binaries that merge within the Hubble time are included.}
   \label{fig:mergetime}
\end{figure}

The merger time distributions for the three different models are shown in
Figure~\ref{fig:mergetime}.  Under all three assumptions about the strength of
kicks, the merger times are predominantly long; that is, greater than $1\,{\rm
Myr}$.  The distributions that we obtain are similar to distributions obtained
by previous work in the field \citep[e.g.~figure~23 of][]{fryer99}.

\subsection{Galactic trajectories}

\begin{figure}
   \includegraphics[width=.72\columnwidth]{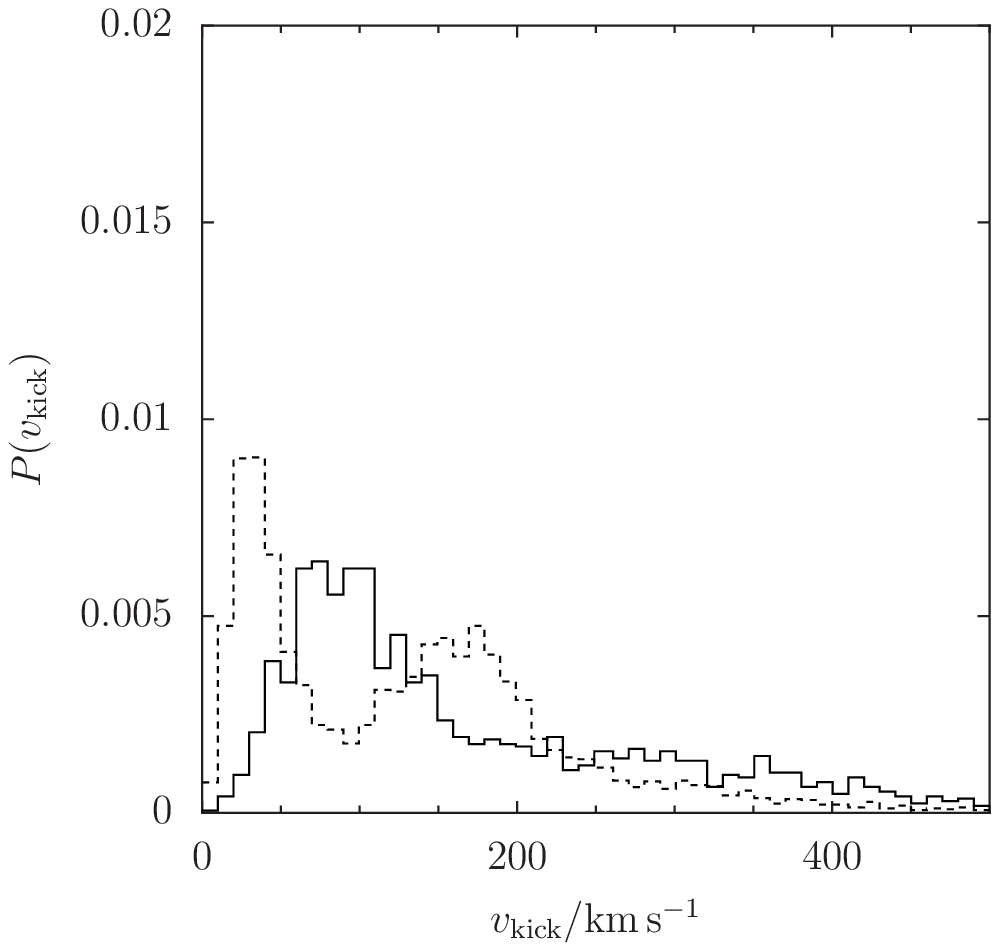}\\
   \includegraphics[width=.72\columnwidth]{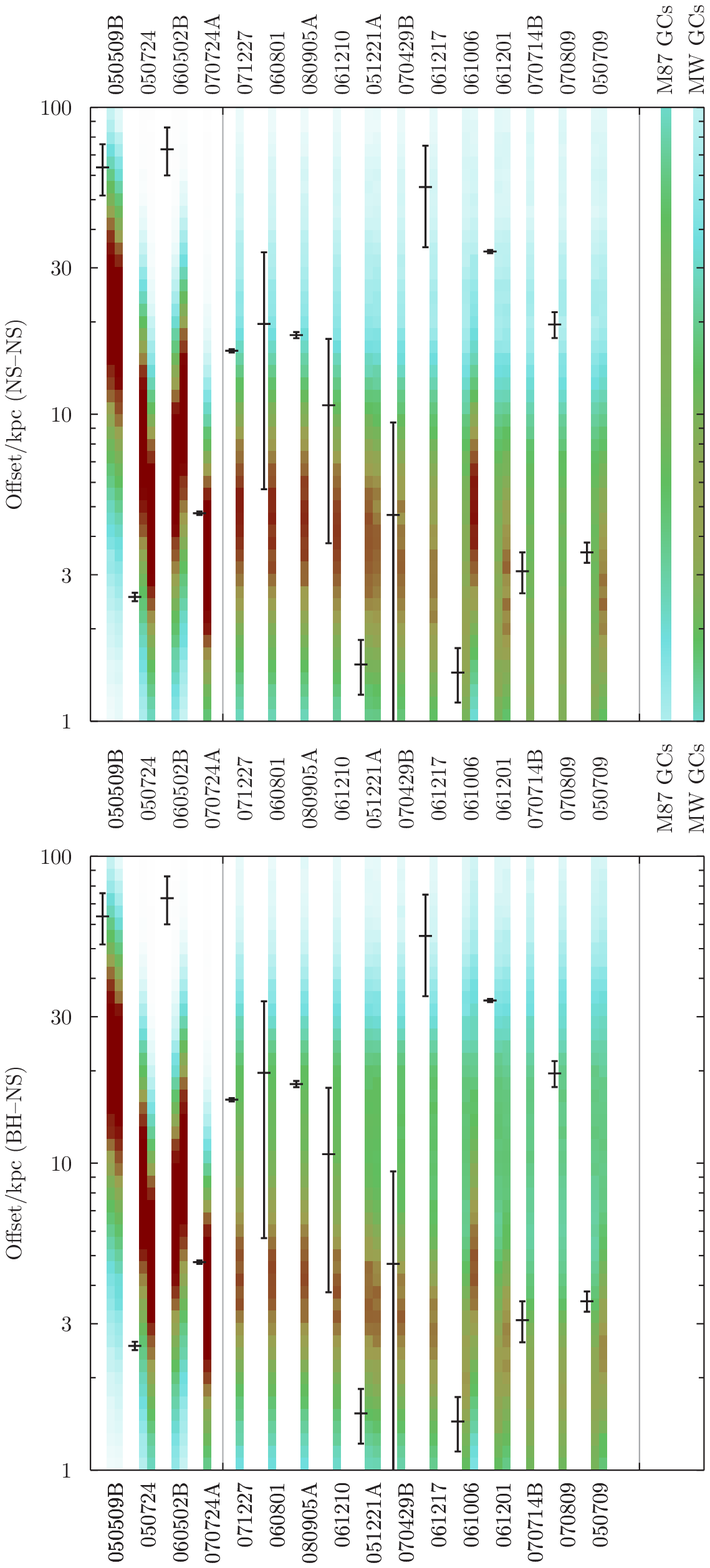}
   \caption{Top: the distributions of birth velocities for NS--NS
   (solid lines) and BH--NS (dashed lines) binaries.  Both first and second
   natal kicks are taken from the ACC02 distribution.  Middle: offset
   distributions of NS--NS binaries around different burst host galaxies
   compared to the observed burst offsets.  The darkness of the colour is
   linearly related to $\ud P(R)/\ud\log R$.  Error bars show one-sigma
   uncertainties in the observed offsets.  Where both observed and predicted
   effective radii are available the additional right-hand stripe shows
   calculations made with the observed $R_{\rm e}$.  The two rightmost
   stripes represent the distribution predicted for the globular cluster systems
   of M87 and the Milky Way.  Bottom: as middle but for BH--NS binaries.
   }
   \label{fig:offsetACCACC}
\end{figure}

\begin{figure}
   \includegraphics[width=.72\columnwidth]{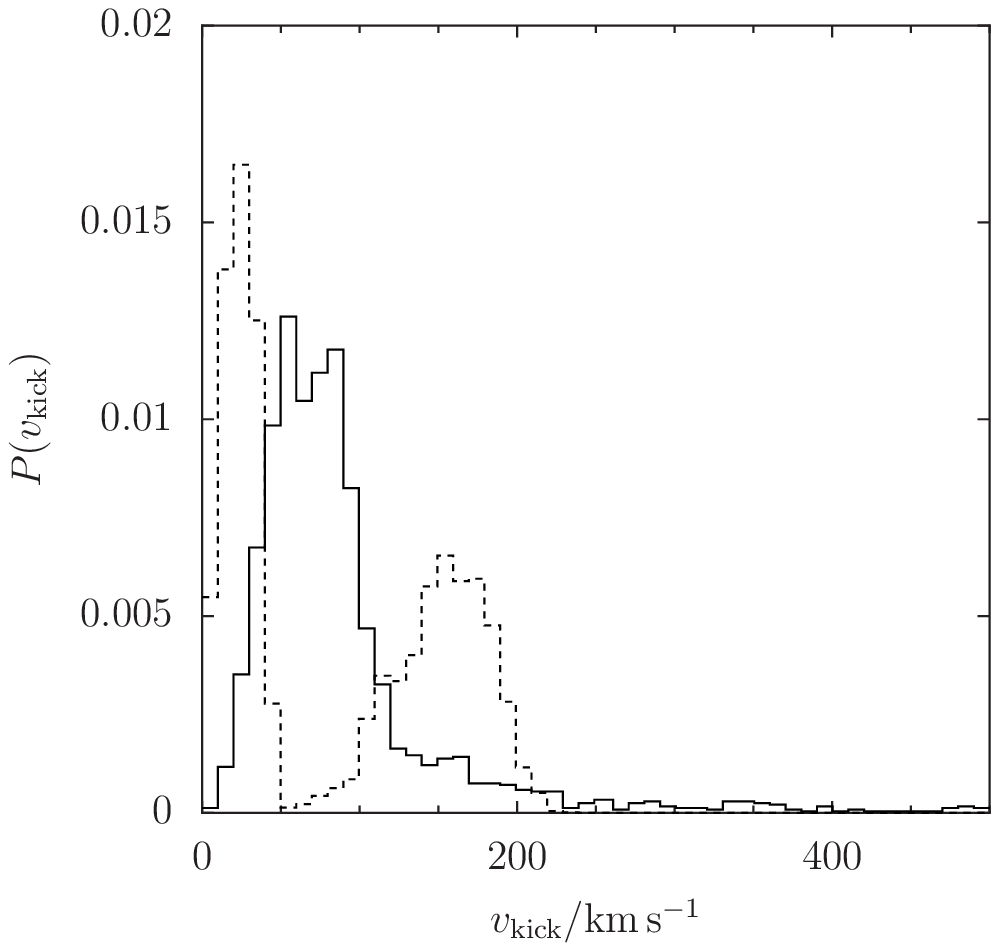}\\
   \includegraphics[width=.72\columnwidth]{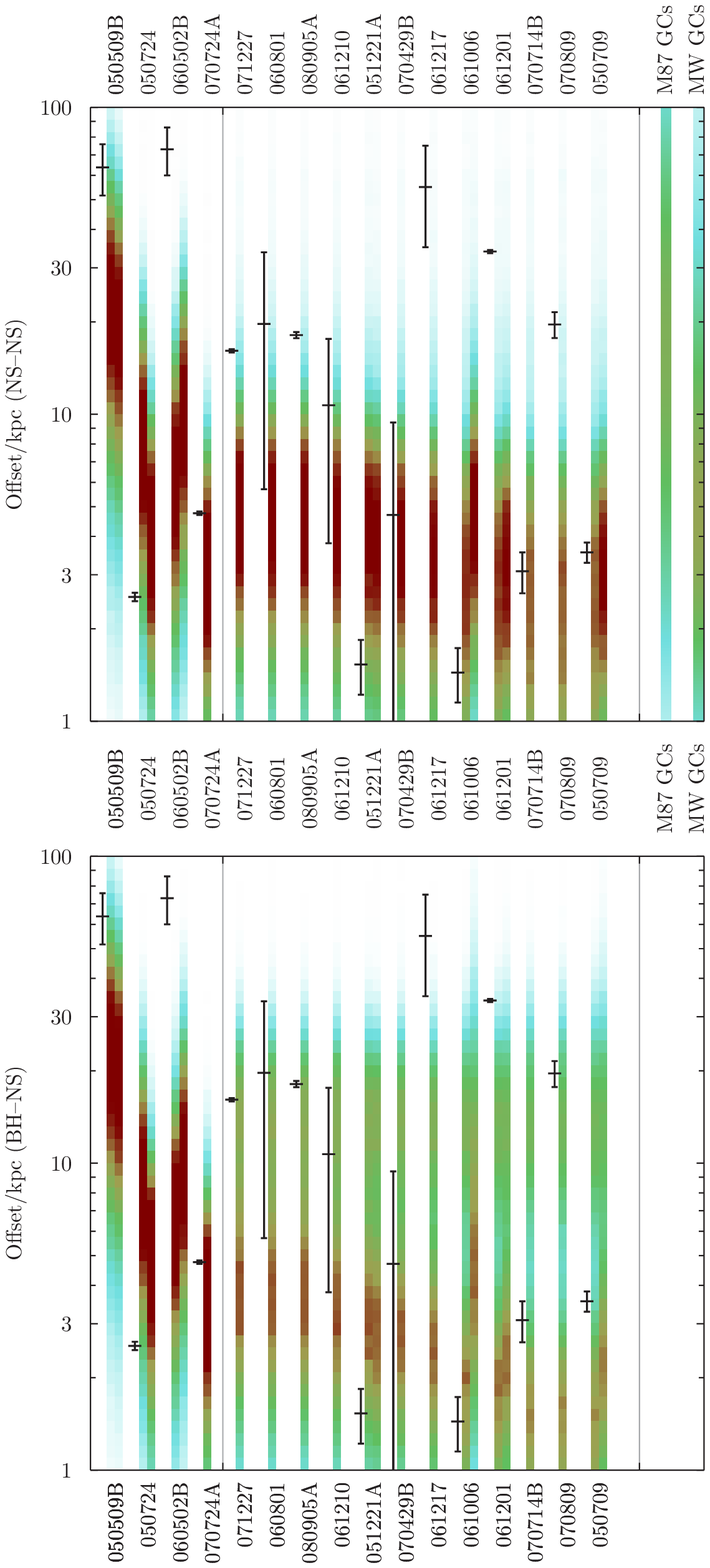}
   \caption{As in Figure~\ref{fig:offsetACCACC}, but the first kick is taken
   from the ACC02 distribution and the second from the DPP05 distribution.
   }
   \label{fig:offsetACCDewi}
\end{figure}

\begin{figure}
   \includegraphics[width=.72\columnwidth]{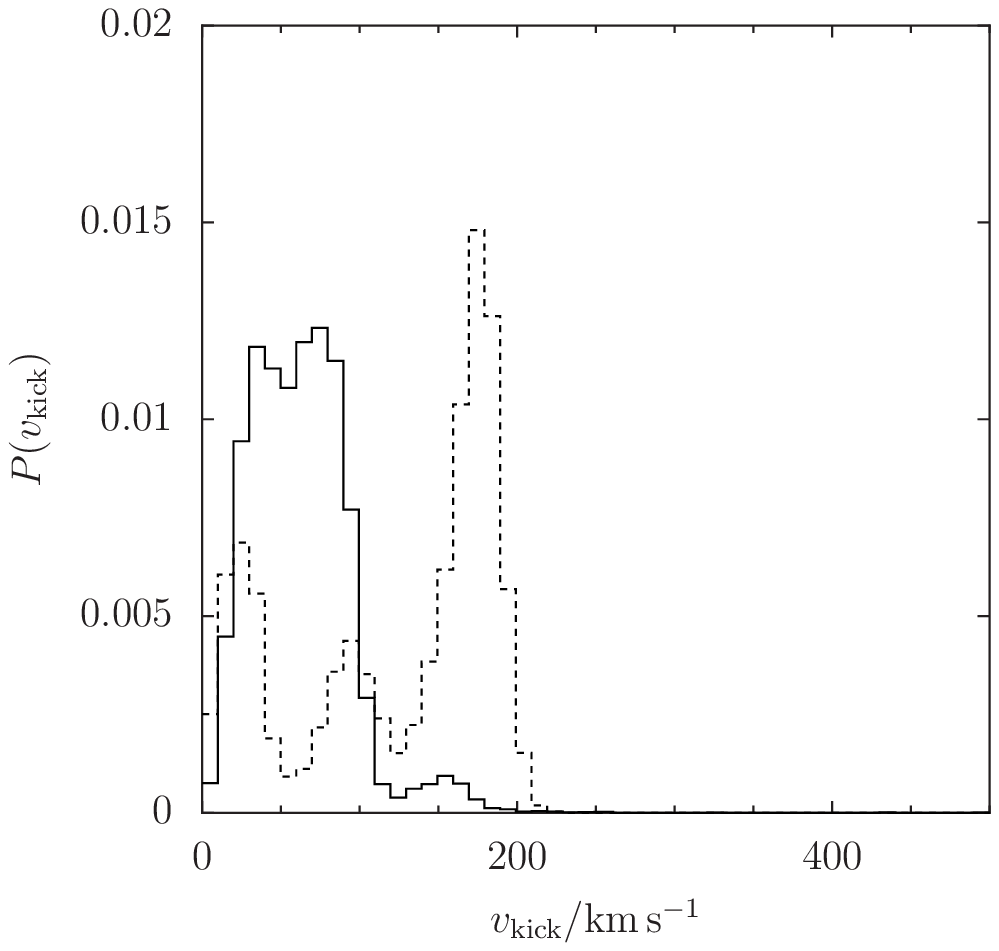}\\
   \includegraphics[width=.72\columnwidth]{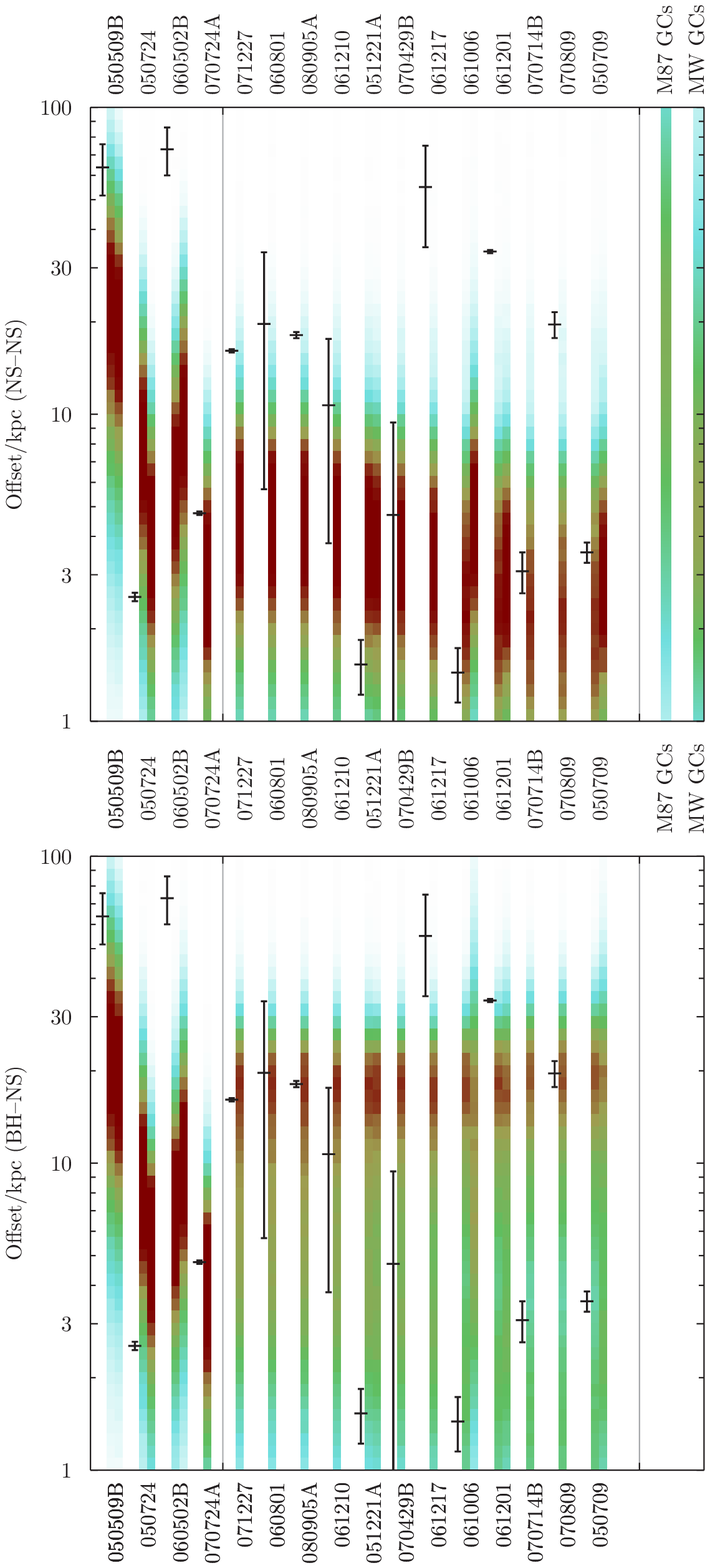}
   \caption{As in Figure~\ref{fig:offsetACCACC}, but with both kicks drawn from
   the DPP05 distribution.
   }
   \label{fig:offsetDoubleDewi}
\end{figure}

In order to predict the offsets of compact binary systems from their host
galaxies at the time of final merger we integrate their motion in the
gravitational potentials of their modelled host galaxies.  For each host we
produce a halo model as described in Section~\ref{sect:burstshosts}; the
parameters of the models are listed in Table~\ref{tab:models}.  We place the
stars initially in an exponential disc and assign the initial location of each
DCO binary with a probability proportional to the disc's luminosity at that
point, i.e. our DCO binaries originate tracing the light of their host galaxies
as is seen for most supernovae \citep{fruchter2006,kelly08}.   
Each DCO binary's initial velocity is set equal to the sum of its circular
velocity in the galactic potential and a kick velocity that originates in the
effects of the supernovae on the binary's centre-of-mass velocity; this is
calculated during the population synthesis.  The orientation of the orbital
plane of the binary with respect to the disc of the host galaxy is taken to be
isotropically distributed.  The binary's trajectory is then calculated
and its position at the time of merger recorded. We view the galaxy from a
direction chosen uniformly across a sphere and record the binary's {\em
projected} offset from the galaxy centre at this time. We compute $10^5$
trajectories for each galactic model in order to obtain distributions of
predicted burst offsets.  These distributions are shown in
Figures~\ref{fig:offsetACCACC}, \ref{fig:offsetACCDewi} and
\ref{fig:offsetDoubleDewi}, in addition to the offsets of observed bursts.

For some of the hosts the inclination of the host disc with respect to our
galaxy is known.  Hence we tested whether the viewing the host from a fixed
angle as opposed to a random angle affected our results.  We found that the
distribution of offsets did not change significantly.

\section{Discussion}

The results of the simulations are presented in Figure~\ref{fig:avse}, which
shows the distribution of binaries in the $a$ vs. $e$ plane, and
Figures~\ref{fig:offsetACCACC} to \ref{fig:offsetDoubleDewi}, which contain the
binary natal velocity distributions and predicted offset distributions.

\subsection{Synthesised binary population}
\label{sect:discuss:pop}
The distributions of compact object binaries in the $a$, $e$ plane that our
population synthesis calculations generate are shown in Figure~\ref{fig:avse}.
A detailed discussion of whether the observed population is consistent with our
prediction requires consideration of the biases in the detection of pulsars and
is beyond the scope of this project, but it can be simply stated that it is
necessary for all of the observed binaries to be able to form under the adopted
description of binary evolution processes.  The $a$, $e$ plots show that, for a
kick distribution where the second kick is weak -- that is, drawn from a
DPP05-like population -- it is not possible to make a NS--NS binary such as
PSR~B1913+16 (the Hulse-Taylor pulsar).  This means that the binary in which it
formed must have experienced a relatively strong second kick.  It does not
necessarily indicate that all NS natal kicks must be strong, as some intricacy
of the evolution may cause some NSs to receive a strong natal kick and some a
weak one, but for the purposes of this study it leads us to prefer a kick
distribution where all the NSs receive a strong kick on formation.  The
$a$--$e$ distribution that we obtain using a strong kick is similar to that of,
for example, \citet{bloom99}, though with an additional population of
long-period binaries.  This difference is not significant for this project as
the long-period population all have merger times greater than the Hubble time.

Examination of the binary birth velocity distributions presented in
Figures~\ref{fig:offsetACCACC} to \ref{fig:offsetDoubleDewi}, which are the
result of the effects of both supernova mass loss and supernova natal kicks on
the space velocity of the compact binaries, shows that there are significant
differences between NS--NS and BH--NS binaries.  In the cases where the second
kick is weak (Figures~\ref{fig:offsetACCDewi}~and~\ref{fig:offsetDoubleDewi}) a
significant population of the BH--NS binaries acquire velocities that are large
relative to the NS natal kick velocities.  These large velocities originate in
the loss of a substantial quantity of mass during the formation of the NS when
the binary is close and hence the stars' velocities large.  The lower compact
companion mass in comparable systems that form NS--NS DCOs limits the amount of
mass that can be lost without the binary breaking up and hence reduces the
effect of this process.  Therefore, for a small second kick the systems with
the largest space velocities are BH--NS binaries.  This is not the case,
however, when the second kick is large (and hence more dominant over the effect
of mass loss); there a much larger proportion of the highest-velocity systems
are NS--NS binaries.

\subsection{Predicted galactic offsets}
The hosts in Figures~\ref{fig:offsetACCACC} to \ref{fig:offsetDoubleDewi} are
sorted in order of decreasing blue-band magnitude.  A generally large scatter
can be seen in their observed offsets, though two of the three most luminous
-- and more massive -- elliptical galaxies host short bursts at typically large
offsets.  These are also typically bursts with shorter durations, which have
been suggested to lie at larger radii from their hosts \citep{troja08}, where
the short durations, and very faint X-ray afterglows would correspond to low ISM
density about the progenitor.  The large scatter observed is consistent with the
models, as shown by the broad probability distributions.

There is a qualitative difference between the behaviour of binaries in the
elliptical galaxies in our sample and those in spiral galaxies.  The elliptical
galaxies are much more massive and have halo circular velocities between 400 and
$700\,\kms$.  These circular velocities exceed the vast majority of the binary
natal velocities regardless of the kick distribution used.  Hence the binaries'
galactic orbits are only mildly perturbed by the kicks and the offset
distributions are determined largely by the effective radii of the hosts.  The
spiral galaxies, on the other hand, have much lower circular velocities of
between 100 and $200\,\kms$.  Hence they have extended offset distributions
given their small sizes, particularly for the BH--NS binaries and the stronger
kick distribution.

\begin{figure}
\begin{center}
GRB 060502B\\
\includegraphics[width=\columnwidth]{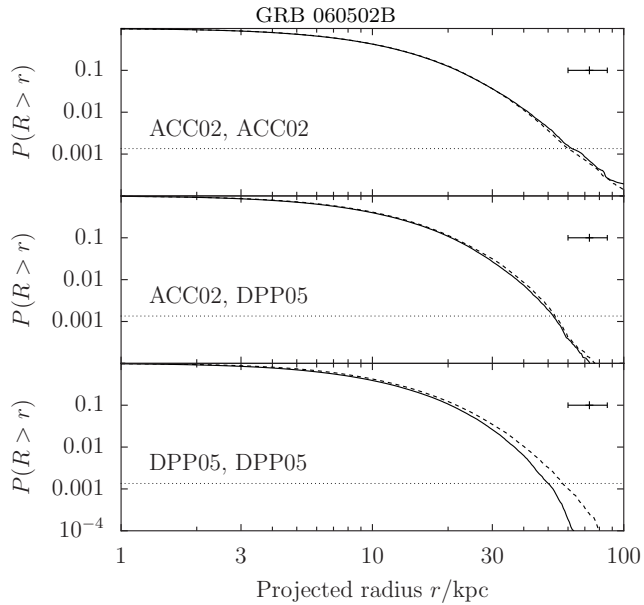}
\end{center}
\caption{The probability of obtaining an offset larger than some projected
radius $r$ as a function of $r$, in our models of the host galaxy of
GRB~060502B.  Solid lines represent the distributions of NS--NS binaries, dashed
lines those of BH--NS binaries.  Distributions have been calculated using the
observed $R_{\rm e}$.  The three panels show the
three different pairs of kick distributions that we considered.  The error bars,
placed at arbitrary heights,
show the one-sigma error on the observed offset.  The dotted line marks the
three-sigma probability.}
\label{fig:cumProb060502B}
\end{figure}

\begin{figure}
\begin{center}
GRB 061217\\
\includegraphics[width=\columnwidth]{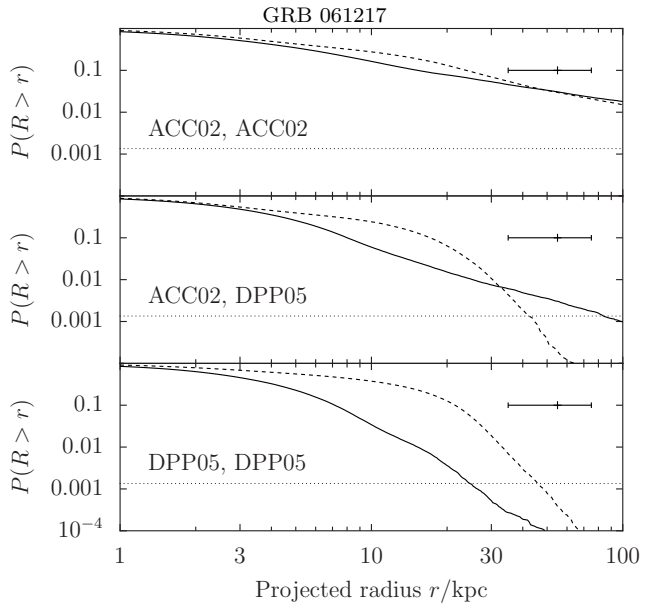}
\end{center}
\caption{As Figure~\ref{fig:cumProb060502B}, but for GRB 061217.  The
calculations use a predicted $R_{\rm e}$ as no observation was available.}
\label{fig:cumProb061217}
\end{figure}

\begin{figure}
\begin{center}
GRB 061201\\
\includegraphics[width=\columnwidth]{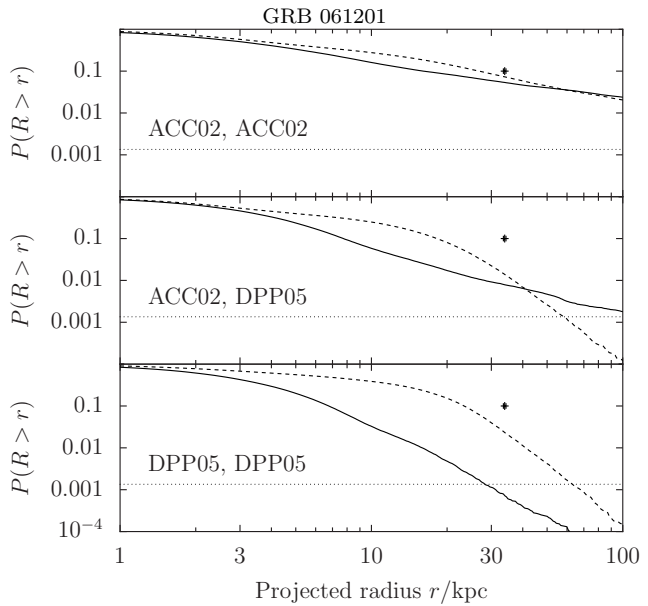}
\end{center}
\caption{As Figure~\ref{fig:cumProb060502B}, but for GRB 061201.}
\label{fig:cumProb061201}
\end{figure}

Our models predict offset distributions that match all of the bursts observed
around spiral galaxies, provided that either the natal kick is relatively
strong, or the progenitors of the bursts with the largest offsets are BH--NS
binaries rather than NS--NS binaries.  The cumulative offset
distributions for the spiral galaxies hosting the two bursts with the largest
offsets are shown in Figures~\ref{fig:cumProb061217} (GRB~061217) and
\ref{fig:cumProb061201} (GRB~061201).  However, the burst with the largest
offset, GRB~060502B which has an elliptical host, is not well fitted by the
models.  The cumulative burst probability, calculated using the measured
$R_{\rm e}$, is shown in Figure~\ref{fig:cumProb060502B}.  Fewer than 0.1\% of
possible hosts lie within its 1-sigma offset position, and fewer than 2\% within
its 3-sigma offset position.  We cannot exclude the model at a 5-sigma level as
the error box then overlaps a large fraction of the host galaxy.

\begin{figure}
\begin{center}
GRB 060502B\\
\includegraphics[width=\columnwidth]{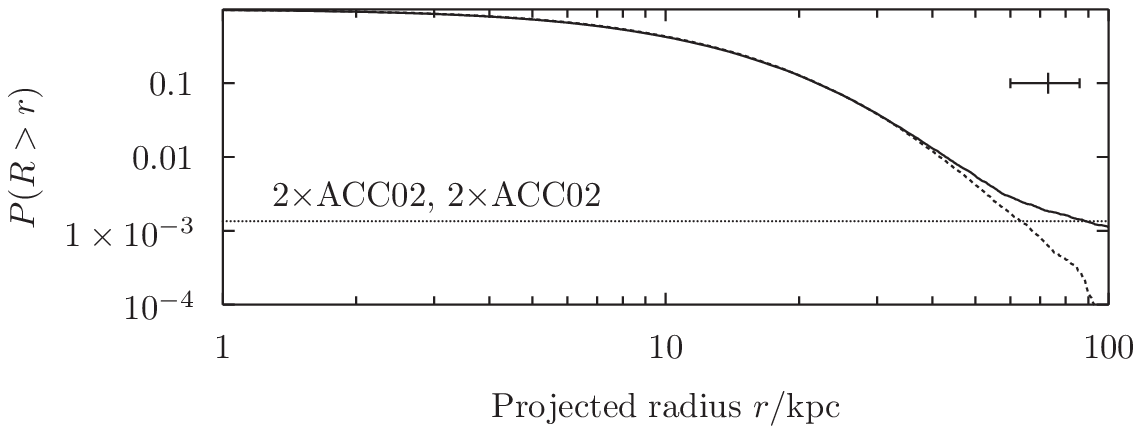}
\end{center}
\caption{As Figure~\ref{fig:cumProb060502B} but showing the effect of doubling
the strength of the ACC02 kick distribution.  The plot shows the probability of
obtaining an offset larger than some projected radius $r$ as a function of $r$,
in our models of the host galaxy of GRB~060502B.  Solid lines represent the
distributions of NS--NS binaries, dashed lines those of BH--NS binaries.
Distributions have been calculated using the observed $R_{\rm e}$.  The three
panels show the three different pairs of kick distributions that we considered.
The error bars, placed at arbitrary heights, show the one-sigma error on the
observed offset.  The dotted line marks the three-sigma probability.}
\label{fig:cumProb.060502B.extra}
\end{figure}

Given the uncertainty in the distribution of natal kicks it is prudent
to investigate the effects of a natal kick stronger than the ACC02 distribution.
Following a suggestion from the referee, we re-calculated the cumulative offset
distribution using an ACC kick distribution scaled up by a factor of two
-- i.e.~where the peak probability was at twice its normal value.  The results
are presented in Figure~\ref{fig:cumProb.060502B.extra}.  The distribution of
offsets from NS--NS mergers remains effectively unchanged as the stronger kicks
merely serve to break up a larger fraction of the binaries.  A small tail of
BH--NS mergers at larger offsets is generated, but it is not significant enough
to explain the burst.

Several other bursts also lie towards the edge of the predicted
distribution.  Only 8\% of predicted burst locations lie outside the measured
offset of GRB~061201, and only 16\% outside the measured offset of GRB~070809
(although in that case the positional error bars are larger).  Furthermore,
there is some uncertainty over the properties of the host of GRB~050509B.
\citet{bloom2006} give its effective radius as $3.47\,{\rm kpc}$, rather than
the $20.98\,{\rm kpc}$ given by \citet{fong09}.  This, in combination with the
velocity dispersion, which \citet{bloom2006} measure to be $250\,{\rm
km\,s^{-1}}$, suggests that it may be more like the host of GRB~060502B
than the model we have used.  If so this
would provide another large offset that would be hard to match with the DCO
model.  These points strengthen the case that there is an excess of large
offsets compared to our predictions.

\subsection{Selection effects and correlations}
\begin{figure}
   \includegraphics[width=\columnwidth]{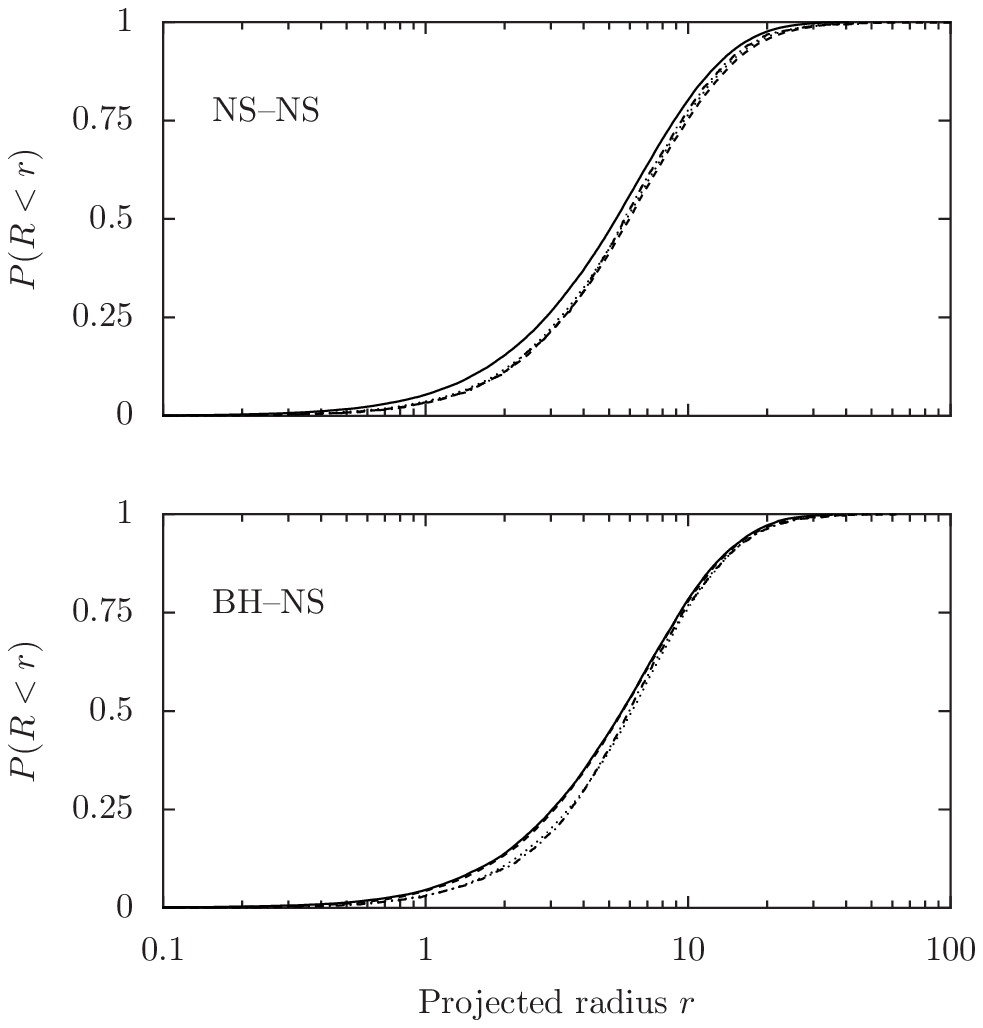}
   \caption{The cumulative probability distribution of projected burst offsets 
   plotted for different times after a starburst.  Solid lines represent bursts
   occurring within $100\,{\rm Myr}$, dashed lines bursts occurring between
   $100\,{\rm Myr}$ and $1\,{\rm Gyr}$ after the starburst, dotted lines bursts
   occurring between $1\,{\rm Gyr}$ and $10\,{\rm Gyr}$ after the starburst and
   dash-dotted lines burst occurring more than $10\,{\rm Gyr}$ after the
   starburst.  The upper panel shows the distributions for NS--NS progenitors,
   the lower panel those for BH--NS progenitors.  The model used is that for the
   host of GRB~060502B.
}
   \label{fig:timevol}
\end{figure}

There are clearly possible selection effects in operation. For example, bursts
expelled large distances from low mass galaxies may be essentially impossible to
associate with their hosts with any confidence. This is the likely explanation
for the apparently host-less GRBs 061201 and 080503 \citep{perley09}.  
A further selection effect worthy of inspection is the possibility that the
majority of DCOs that form in elliptical galaxies have merged before
we observe their hosts.  If the DCOs that merge at small offsets come
predominantly from the population with short merger times then we would not
expect to see them in nearby, old giant ellipticals.  To investigate this we
have plotted the distribution of projected burst offsets for our model of the
host of GRB~060502B as a function of time since a single star formation burst
(Figure~\ref{fig:timevol}).  The distributions for different times are
essentially identical, as the galaxy rotation velocities are much larger than
the supernova kick velocities.  Furthermore, the fraction of SGRBs
observationally associated with old, rather than young populations is at least
$\sim 25\%$, based on the fraction of identified hosts that are elliptical
galaxies.  This strongly suggests that the rapidly merging population cannot
dominate at the current epoch.

Other correlations of burst properties with those of their hosts have been
reported elsewhere.  In particular, \citet{troja08} suggest that bursts with
extended emission typically form at smaller projected distances from their hosts
than very short bursts. They suggest that this may be due to the bursts with
extended emission occurring via NS-BH mergers, which may travel less far.
Our population synthesis indicates that this is not the case, and that BH--NS
and NS--NS populations have broadly similar radial distributions, with the
BH--NS population being {\it more} extended as explained in
Section~\ref{sect:discuss:pop}. For the preferred strong kick we find that the
distributions are almost identical (see Figure~\ref{fig:offsetACCACC}).

A second reported correlation was that of \citet{rhoads08}, who showed that
there is an anticorrelation between the absolute magnitude of the host, and the
isotropic energy release of the burst. Our results confirm this, however, in
the light of previous discussion it is possible to recast this correlation in
terms of the observed offsets from host galaxies, notably, that the offset and
energy appear to be anti-correlated as well. It is therefore interesting to
investigate if the source of the correlation is the due physically to the
difference in host absolute magnitude, the offset, or another physical
mechanism which can explain both. 

There is a correlation between the direction of the natal velocity of a DCO
binary and the orientation of its orbital plane.  In particular, if the natal
kick is weak and hence the binary natal velocity is dominated by the loss of
mass in the second supernova explosion, then the natal velocity should be
perpendicular to the orbital axis.  If the gamma-ray emission is beamed along
the orbital axis then this correlation has the potential to increase the
magnitude of the {\it observed} offsets.  This effect was considered at an
early stage in these calculations and found to be small once realistic binary
evolution and galactic trajectories were taken into account.

\subsection{Host galaxy selection}
An ongoing problem with studies of SGRBs is the association of a given burst
with its host. In a few cases optical afterglow positions provide high
confidence associations showing that a burst lies on the stellar field of a
given host. However, in many cases this was not possible, and different
approaches have been taken. For example, in the case of GRB 060502B,
\cite{bloom2007} identify the host as the giant elliptical considered here with
a large offset, whilst \cite{bergerhiz} adopt a strategy of identifying the
brightest extended object {\em within} the XRT error box, which leads to a
different, fainter host at a smaller offset.  This problem is most severe for
GRBs 050509B and 060502B.  The probability of  random association with a
background galaxy some distance from the burst position is frequently rather
similar to the probability of a chance alignment within the error circle
\citep{levan060912A}.  In such circumstances it becomes very difficult to
identify the host with high confidence, and it is likely that strong
constraints will only come with the build up of a larger sample, in which it is
possible to say if, for example, large elliptical galaxies are
over-represented. 

Of course, in part the difficulty in locating burst afterglows may be due to the
progenitors of the bursts themselves. If NS-NS binaries are expelled from their
hosts with large velocities then their optical afterglow brightness will be
significantly suppressed due to the low ISM density, since the afterglows are
caused by external shocks. In very low density media the cooling break may even
move above the X-ray band, and result in very faint afterglows. Indeed,
\citet{troja08} have suggested that bursts with shorter durations, and fainter
X-ray afterglows lie systematically at larger distances from their hosts than
the longer duration brighter bursts. 

A recent and closely-related paper \citep{2010arXiv1007.0003B} has
discussed the possible origin of short GRBs with optical afterglows with
significant offsets.  They concentrate on bursts with optical afterglows as
having smaller positional errors, and show that for all but one burst where the
identity of the host is in doubt the lowest probabilities of chance co-incidence
is associated with bright galaxies at offsets of a few tens of kpc.  Their
results are in agreement with ours, in that they state that the observations fit
a model where these offsets arise naturally in merging NS--NS binaries that have
been kicked out of their host galaxies.  They do not find any additional
contribution at larger offsets necessary; on the other hand they do not carry
out host-by-host modelling as we do and they neglect the bursts with only X-ray
detections that have the largest offsets.

\section{Globular clusters}

An alternative explanation of the very large offsets of some SGRBs from their
host galaxies is that they occur in compact binaries residing within the
globular cluster systems of their host galaxies.  Within the dense environments
at the cores of globular clusters NS--NS binaries can form through dynamical
interactions \citep{davies1995}.  Such a scenario for the formation of SGRBs was
proposed by \citet{grindlay2005}, who show that, based on some simple
assumptions about the formation channels of NS--NS binaries within GCs and
plausible number densities of neutron stars and low-mass X-ray binaries, a
significant fraction of SGRBs could form via this channel.

\subsection{Radial distribution}

\begin{figure}
\begin{center}
\includegraphics[width=\columnwidth]{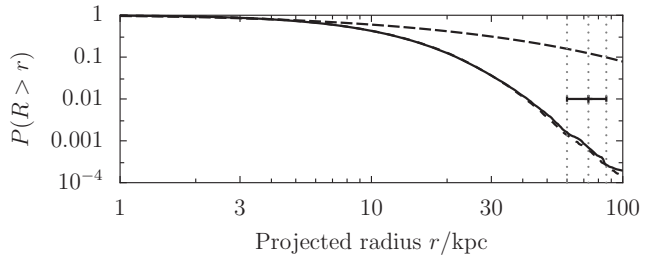}
\end{center}
\caption{The long-dashed thick line shows the offset distribution of SGRBs,
calculated under the assumption that they occur via the merger of
dynamically-generated NS--NS binaries in the cores of globular clusters and
hence follow their projected spatial distribution.  The globular cluster
population chosen is that of M87.  The error bar, placed at arbitrary height,
represents the offset of GRB~060502B from its putative massive elliptical host
galaxy.  The thin solid and short-dashed lines are the results from the standard
model of a field population of DCO binaries as presented in
Figure~\ref{fig:cumProb060502B}.  The thin solid line shows NS--NS binaries and
the thin dashed line BH--NS binaries.}
\label{fig:cumProbM87}
\end{figure}

To assess the viability of this alternative, we have plotted the galactic
distribution of gamma-ray bursts assuming that they follow the globular cluster
population of two galaxies; M87 and the Milky Way.  M87 is a giant elliptical
galaxy in the Virgo cluster, with a population of roughly 14000 globular
clusters \citep{harris09}.  The distribution of its clusters has a very large
spatial extent, extending to at least 100\,kpc.  The Milky Way has a more modest
distribution of about 160 clusters with the most distant also being at roughly
100\,kpc \citep{djorgovski94}.  We plot the projected radial distributions of
the two globular cluster systems in Figures~\ref{fig:offsetACCACC} to
\ref{fig:offsetDoubleDewi}.  For M87 we utilise the fits given by
\citet{harris09}, whereas for the Milky Way we take the fit to the 3-dimensional
number density distribution provided by \citet{djorgovski94} and project it.
For simplicity we assume that the dynamical formation of the NS--NS binaries
which may lead to SGRBs proceeds in a similar manner across all the GCs.
Hence we obtain projected offset distributions, which are plotted on the
right-hand side of Figures~\ref{fig:offsetACCACC} to
\ref{fig:offsetDoubleDewi}, along with the cumulative probability distribution
in Figure~\ref{fig:cumProbM87}.  Both distributions are much more extended than
the kick-derived offset distributions, and are consistent with the most-offset
bursts.  

\subsection{Formation rates}

To assess the viability of globular cluster systems as formation sites for SGRB
progenitors it is instructive to estimate the formation rates within the
clusters and galaxies.  To simulate the populations here we evolved
$4\times10^7$ binaries, which when using strong (ACC02) kicks produced 1659
NS--NS binaries and 4253 BH--NS binaries, counting only those binaries that
merge within $1.4\times10^{10}$ years.  Once we have corrected for the fact that we only
sample the heavy end of the IMF this leads to specific merger rates of
$15\,f_{\rm b}$ and $38\,f_{\rm b}\,(10^9\,\Mo)^{-1}\,t_{\rm Hubble}^{-1}$.
Here $f_{\rm b}$ is the fraction of stellar mass in binary systems.  For
globular clusters we extrapolate from the observation of a single
dynamically-formed NS--NS binary in a Galactic globular cluster, M15-C, with a
merger time of roughly 300\,Myr, amongst the roughly 150 Galactic globular
clusters.  Utilising the observations of \citet{Rhode05} which suggest that the
number of globular clusters per $10^9\,\Mo$ of stellar mass, $T$, lies between 1
(for field spiral galaxies) and 4 (for cluster elliptical galaxies) gives a
NS--NS merger rate from globular clusters of $0.31 T\,(10^9\,\Mo)^{-1}\,t_{\rm
Hubble}^{-1}$.  This is obviously much smaller than the field merger rate even
for elliptical galaxies but there is a severe underestimate for several reasons.
Pulsar surveys of galactic globular clusters are incomplete, and the inclusion
of only a single system in our calculations samples only a small range of
possible merger times.  Based on simple scattering simulations of binaries in
globular clusters \citet{grindlay2005} show that this simple extrapolation is an
underestimate by a factor of between 10 and 100, which brings the rate estimate
to a similar magnitude to that from field binaries.  Furthermore, the massive
elliptical galaxy M87 has a value of $T\simeq8$ for the blue globular cluster
sequence alone \citep{Brodie06}.  Clearly more detailed simulations of globular
clusters are required to form a more accurate estimate of their production rates
of NS--NS binaries, but this crude analysis shows that the rates could be
comparable, and that given a sample of nearly 20 SGRBs it would not be
surprising for the progenitor binary of one or more of them to have formed in a
globular cluster.  Furthermore, the burst that we suggest is most likely to
have formed in a globular cluster, GRB~060502B, has a giant elliptical host
which would be expected to have a larger specific globular cluster fraction.

\section{Summary}

We have synthesised populations of NS--NS and BH--NS binaries and used them to
predict the galactic offset distributions of short gamma-ray bursts, given the
observed properties of their host galaxies.   The offsets of all but one short
gamma-ray burst are found to be consistent with this picture.  However, the
burst GRB~060502B apparently has an inconsistently large offset.  It is
plausible that this inconsistency results from a mis-identification of the host
galaxy, but we show that the large offset is consistent with the merger of
a NS--NS binary that has been created by dynamical encounters within a globular
cluster.  The offset distribution arising from the globular cluster systems of
the Milky Way and a local massive elliptical galaxy, M87, are consistent with
the observed offsets of all short GRBs, including GRB~060502B, and that the
production rate of such binaries may be comparable with that of field DCO
binaries. 

\section*{Acknowledgements}
The authors would like to thank the anonymous referee for their constructive
comments, which helped to improve this paper.  RPC thanks the Wenner-Gren
Foundation for support. This work was supported by the Swedish Research Council
(grant 2008-4089).  The calculations presented in this paper were carried out
using computer hardware purchased with grants from the Royal Physiographic
Society of Lund.

\bibliographystyle{mn2e.bst}
\bibliography{new.bib}

\label{lastpage}

\end{document}